\documentstyle[epsfig]{ioplppt}                

\begin{document}
\jl{4}

\title{ Strangeness and Pion Production as Signals of QCD Phase Transition
}

\author{Marek Ga\'zdzicki\ftnote{3}{
e--mail: marek@ikf.uni--frankfurt.de}}

\address{Institut f\"ur Kernphysik, Universit\"at Frankfurt \\ 
August--Euler Str. 6, D--60486 Frankfurt, Germany}

\begin{abstract}
A systematic analysis of data on strangeness and pion
production in nucleon--nucleon and central nucleus--nucleus
collisions is presented.
It is shown that at all collision energies the pion/baryon and
strangeness/pion ratios indicate saturation with the
size of the colliding nuclei.
The energy dependence of the saturation level suggests that
the transition to the Quark Gluon Plasma occurs
between 15 A$\cdot$GeV/c
(BNL AGS) and 160 A$\cdot$GeV/c (CERN SPS) collision energies. 
The experimental results interpreted in 
the framework of a  statistical approach
show that
the effective number of degrees of freedom increases  
in the course of the phase transition
and that the  plasma created at CERN SPS energies 
may have a temperature
of about 280 MeV (energy density $\approx$ 10 GeV/fm$^{-3}$).
The presence of the phase transition can lead to the
non--monotonic collision energy dependence of the strangeness/pion ratio.
After an initial increase the ratio should drop to the characteristic value
for the QGP.
Above the transition region the ratio is expected to
be collision energy independent.
Experimental studies of central Pb+Pb collisions in the 
energy range 20--160 A$\cdot$GeV/c  are urgently 
needed in order to localize the threshold energy,
and study the properties of the QCD phase transition.
\end{abstract}

%
%

\vspace{1cm}
{\it Talk given at International Symposium on `Strangeness in Quark
Matter 1997', Thera (Santorini), Greece, April 14--18, 1997.}

\newpage

\section{Introduction}

In this paper we review the recent status of our search for a QCD 
phase transition
to the Quark Gluon Plasma \cite{qgp} by a systematic analysis of
strangeness and pion (entropy) production in nuclear collisions.
There are important reasons to select  strangeness \cite{Ko:86} and 
entropy \cite{Va:82} 
as basic observables.
Both are defined in any form of strongly interacting matter.
Their equilibrium values are directly sensitive to the basic 
properties of  matter: effective number of degrees of freedom and
their effective masses.
Entropy and strangeness production are  believed to be produced at the
early stage of the  evolution of a system  created in nuclear 
collisions and therefore they can allow us to `measure' properties
of matter at very high energy densities.

Our strategy of data\footnote{ Data obtained by about 100
different experiments are used. The references to the original
experimental works can be found in the compilation papers 
\cite{Ga:1,Ga:2,Ga:3}.} 
analysis \cite{Ga:1,Ga:2,Ga:3,Ga:4,Ga:5,Ga:6} reviewed here
can be summarized as follows:\\
1. We study the dependence of the properly normalized entropy
(mainly determined by pion multiplicity) and
strangeness (mainly determined by kaon and hyperon yields)
production on the volume of the colliding nuclear matter at a
fixed collision energy.
We demonstrate that a fast saturation occurs.    
The simplest qualitative interpretation is, that equilibration of
entropy and strangeness takes place (Section 2). \\
2. We study the dependence of the saturation levels on the collision
energy.
We demonstrate that the saturation levels for both entropy and strangeness
show an unusual change between AGS ($\approx$15 A$\cdot$GeV/c) and SPS
($\approx$160 A$\cdot$GeV/c) energies.
We interpret this effect as due to the localization
of the threshold energy for the QGP creation in the above energy
range (Section 3). \\
3. Finally, we formulate a simple statistical model for entropy and
strangeness production in nuclear collisions and show that
the experimental results at SPS energes can be quantitatively described
assuming creation of a QGP (Section 4).
A critical discussion of the basic assumptions used in the model
is given at the end of this section.

\section{Volume Dependence}

Experimental data on pion\footnote{Here we use pion multiplicity instead
of entropy in order to start the analysis from `raw' experimental data.
An improved experimental
estimate of entropy is given in the next section.}
 and strangeness production in central
nucleus--nucleus (A+A) and all inelastic nucleon--nucleon (N+N)
collisions are shown in Fig. 1  as a function of the number
of participant nucleons, $\langle N_P \rangle$, for various
collision energies.
In order to eliminate a trivial volume dependence, the normalized
multiplicities are studied:
\begin{equation}
\frac {\langle \pi \rangle} {\langle N_P \rangle}
\end{equation}
and
\begin{equation}
E_S \equiv
\frac{ \langle \Lambda \rangle + \langle K 
+ \overline{K} \rangle }  {\langle \pi \rangle},
\end{equation}
where $\langle \pi \rangle$, 
$\langle \Lambda \rangle$, and
$\langle K+\overline{K} \rangle$ are the
mean multiplicities of pions, $\Lambda(+\Sigma^0)$ hyperons, and
kaons and antikaons, respectively.
For all energies a similar behaviour is observed:
a rapid change between results for N+N interactions and  intermediate mass
nuclei ($\langle N_P \rangle \approx$ 50) is followed by a well defined
region in which the 
normalized pion and strangeness production is almost constant.
We interpret the observed saturation as a result of an equilibration of entropy
and strangeness yields.
As the production rates are steeply decreasing functions of the temperature
the equilibration can be expected to happen at the early stage of the 
collision. 
Thus the measured equilibrium values may reflect the properties of the
initially created matter.

\section{Energy Depenedence}

The collision energy dependence of the normalized entropy and strangeness
production is shown in Fig. 2.
The energy dependence is studied using the Fermi energy
variable \cite{Fe:50,La:53}:
\begin{equation}
F \equiv \frac {(\sqrt{s}_{NN} - 2 m_N)^{3/4}}  {\sqrt{s}_{NN}^{1/4}},
\end{equation}
where $\sqrt{s}_{NN}$ is the c.m. energy for a nucleon--nucleon pair and
$m_N$ is the nucleon mass.
There are several advantages in using $F$ as an energy variable.
The measured mean pion multiplicity in N+N interactions 
is approximately proportional
to $F$ \cite{Go:89,Ga:4}.
In the Landau model \cite{La:53} both the entropy and
the  temperature of the initially created matter (for
$\sqrt{s}_{NN} \gg 2 m_N$) are also proportional to $F$.

The `entropy' presented in Fig. 2 is calculated as:
\begin{equation}
S \equiv \langle \pi \rangle + \kappa \langle K+\overline{K} \rangle +
                               \alpha \langle N_P \rangle,
\end{equation}
where the two last components take into account kaon production and pion
absorption \cite{Ga:4,Ga:5}.
Thus $S$ can be treated as the 
inelastic entropy measured in  pion entropy
units.

The normalized `entropy' for central A+A
collisions  at low energies (AGS and below) follows the dependence 
established by N+N data.
It is about 30\% higher for A+A collisions at SPS energies 
(the data of NA35 and NA49 Collab.) 
than  for N+N interactions. 

The  energy dependence of the $E_S$ ratio is also shown in Fig. 2.
The results for N+N interactions are scaled 
by a factor of 3.6 to fit A+A data at AGS
for a better comparison of the energy dependence. 
A monotonic increase of the $E_S$ ratio between Dubna energy 
(p$_{LAB}$ = 4.5 A$\cdot$GeV/c) and SPS energies 
is observed.
In the range from AGS to SPS  the $E_S$ ratio
for N+N interactions is enhanced by a factor of about 2.
A qualitatively different  dependence  is
seen for central A+A collisions.
An increase of  $E_S$ between Dubna\footnote{
Note that the saturation of the $E_S$ with $\langle N_P \rangle$
at Dubna energy is still not established experimentaly, see Fig. 1}
 and AGS energies
is followed by a weak (if any) change of  $E_S$ between AGS and  SPS
collision energies.

Let us now try to understand the observed energy dependence on a
qualitative level.
In the generalized Landau model \cite{Ga:4} the inelastic entropy
is given by:
\begin{equation}
S \sim  g^{1/4} \langle N_P \rangle F,
\end{equation}
where $g$ is the effective number of degrees of freedom.
Thus the observed deviation of the data for A+A collisions 
from the Landau scaling, $S/\langle N_P \rangle \sim F$,
can be interprated as due to an increase of the effective number of 
degrees of freedom when crossing the transition collision energy. 
The magnitude of this increase can be estimated,
within the model, as the fourth power of the ratio of 
slopes of straight lines describing  
low and high energy A+A data:
1.33$^4$ $\approx$ 3 \cite{Ga:4}.
Note that this estimation can be treated as an upper limit as it
is based on the assumption that the inelasticity is the same in the
N+N and A+A collisions.

The second dominant effect of the transition 
to a QGP is the reduction of the effective masses of degrees of
freedom.
Basic thermodynamics tells us that for massless particles 
the ratio {\it (particle number)/entropy}  is independent of the temperature.
For massive particles the ratio increases with $T$ at low
temperature and approaches the  saturation level (equal to the 
corresponding ratio for massless particles) at high temperatures,
$T \gg m$.
This property can be used to estimate the magnitude of the 
effective mass
of strangeness carriers in strongly interacting matter.
The $E_S$ ratio is approximately proportional to the ratio 
{ \it (number of strangeness carriers)/entropy (strangeness/entropy)} 
and therefore
its temperature (collision energy, $F$) dependence should be sensitive
to the effective mass of strangeness carriers. 
Reducing  the mass of
strangeness carriers should cause a weaker dependence of
the $E_S$ ratio on the collision energy.   
An increase of the $E_S$ ratio in the energy 
range of $ F < $ 2 GeV$^{1/2}$ 
can be interpreted as due to the large effective mass of strangeness
carriers
(kaons or constituent strange quarks, $ m_S \approx $ 500 MeV/c$^2$ )
in comparison to the temperature of matter, $T < T_C \approx $ 150 MeV.
At temperatures above $T_C$, the matter is in the form of a QGP
and the mass of strangeness carriers is equal to the mass of
current strange quarks, $ m_S \approx $ 150 MeV/c$^2$, consequently
$ m_S \leq T $.
Thus a much weaker dependence of the $E_S$ ratio on $F$ is expected 
in the high energy region where the creation of the QGP takes place.
The equilibrium value
of the {\it strangeness/entropy} ratio
is higher in hadronic matter (HM) than in the QGP at very high temperatures
\cite{Ka:86}.
This is due to the fact that it is proportional
to the ratio of the effective number of strangeness degrees of freedom
to the number of all degrees of freedom.
This ratio is lower in a QGP than in HM.
At low temperature, however, the {\it strangeness/entropy}
ratio is lower in HM than in a QGP.
This is caused, as previously discussed, by the high mass of strangeness
carriers in HM. 
Thus, in general, a transition to the QGP may lead to an increase or a decrease
of the {\it strangeness/entropy} ratio depending at which temperatures
of the QGP and the HM the  comparison is made.

The presented data suggest that the transition is associated with a
decrease of the {\it strangeness/entropy} ratio.
Thus one can expect a non--monotonic energy dependence of the 
{\it strangeness/pion} ratio; an initial increase of the ratio 
should be followed by a drop of this ratio to the  characteristic value
for the QGP. Above the transition region the ratio is expected to be
collision energy independent.

\section{Model of QGP in A+A}

Encouraged by the qualitative agreement of the data with the  hypothesis
of the equilibrium QGP creation in the early stage of A+A collisions at SPS, 
we attempt to make a quantitative comparison using the simplest version
of the generalized Landau model \cite{Ga:4,Ga:6}.
In the first part of this  section we formulate the model and
compare the results with the experimental data.
In the second part we critically review the basic assumption made.

\subsection{Model Formulation and Results}

We assume that inelastic energy  (energy carried by the produced particles)
is deposited and thermalized 
in a volume equal to the
Lorentz contracted  volume of 
the two overlapping nuclei
(we only consider collisions of two
identical nuclei only). In this volume an equilibrated QGP is formed.
For  simplicity, we assume that the 
two barionic fluids containing  baryons of projectile and target
nuclei are initially decoupled from the baryon--free QGP.
Furthermore,
we assume that the inelastic entropy\footnote{
Strictly speaking a fraction (less than 5\% at CERN SPS) of the
inelastic entropy is absorbed by baryonic fluids  in the later stages
of the system evolution. A correction for this effect is applied, see Eq. 4.}
and strangeness are not changed
during  the system evolution.

The inelastic energy at SPS was measured to be (67$\pm$7)\% of
the available energy \cite{Ba:94}, the same for S+S and Pb+Pb collisions.
It is also weakly dependent on the collision energy  between 
AGS and SPS \cite{St:96}.
In the model we use an approximation of the uniform distribution of
matter in the colliding nuclei. The radius of such an effective
nucleus is calculated as:
\begin{equation}
R = r_0 \cdot A^{1/3} = (\frac {3 A} {4 \pi \overline{\rho}})^{1/3},
\end{equation}
where $A$ is a nuclear mass number and $\overline{\rho}$ is an
average nuclear density calculated using a parametrization of the nuclear
density distribution as given in Ref. \cite{Da:85}.
The resulting values of $r_0$ are 1.34 fm and 1.27 fm for S and Pb nuclei,
respectively.
The value of $r_0$ = 1.3 fm is used in the calculations.
Similar $r_0$ values are obtained from the fits to the A+A inelastic
cross section \cite{cs} using a hard sphere approximation.

The QGP is assumed to consist of massless gluons and massive $u, d$, and $s$
quarks, and the corresponding antiquarks.
The average mass of light quarks was taken to be 7.2 MeV \cite{Le:96}.
The strange quark mass was taken to be 175$\pm$25 MeV \cite{Le:96}, 
this mass is determined at an energy scale of 1 GeV. 
The conversion factor between the calculated entropy and the `entropy'
evaluated from the experimental data is taken to be 4
(entropy per pion at T $\approx$ 150 MeV).
The conversion factor between total strangeness and strangeness measured
by the sum of $\Lambda$ and $K+\overline{K}$ yields is taken to be
1.36 according to the N+N data and a procedure developed in \cite{Bi:92}.
It should be stressed that the model formulated in the above way
has   no free parameters.

The resulting production of entropy and strangeness in the energy range
30--500 A$\cdot$GeV is represented by dashed lines in Fig. 2.
The agreement with the data is surprisingly good.
The analysis suggests that plasma created at the SPS  has an energy 
density of about 10 GeV/fm$^3$ and a temperature of about 280 MeV.

\subsection{Discussion of Basic Assumptions}

In the following we review basic model assumptions concerning
the early stage volume in which thermalization takes place,
the thermalized energy and the production of entropy and strangeness.

{\bf Early Stage Volume.} 
It was Fermi \cite{Fe:50} who for the first time introduced
the Lorentz contracted initial volume of two overlapping protons
as an early stage volume in which thermalization of the
available energy occurs.
This assumption was later taken by Landau and collaborators 
\cite{La:53}
in order to define initial conditions for the hydrodynamical expansion
of matter created in A+A collisions.
This volume can be treated as a maximum
volume (no compression of colliding matter) in which all incomming matter
has a chance to be excited.
For central collisions of nuclei as large as the S--nucleus 
and energies as high as CERN SPS energies the above volume
is large enough to use a grand canonical approximation for entropy and
strangeness production \cite{Ra:80}.
The volume is also large enough to be calculated from the initial
geometrical volume of the nucleus (here the limitation is given by the
longitudinal dimension).
The effect of `smearing' due to the uncertainty principle can be neglected.

{\bf Thermalized Energy.}
Fermi \cite{Fe:50} and Landau \cite{La:53} assumed that the full
available energy in the c.m. system is thermalized in the early stage
volume.
Instantenous decay of this matter into final state hadrons was assumed
by Fermi.    
Landau, following Pomeranchuk's  suggestion \cite{Po:51}, assumed that 
the matter before freeze--out undergoes a hydrodynamical expansion.
Both models are in direct contradiction with the data.
They predict that the rapidity distribution of nucleons is narrower than
the  one for pions
(due to the mass difference) in contrary to the measured distributions
in p+p and A+A collisions \cite{Ba:94}.
In addition they predict similar\footnote{
Exact results depend on the freeze--out conditions assumed
in the model.}
rapidity distributions of baryons and antibaryons, but the data for p+p and
A+A collisions show strong differences between $\Lambda$ and 
$\overline{\Lambda}$ distributions \cite{Ga:91,Al:94}.

Based on this observation and guided by the partonic structure of the
nucleon  Pokorski and Van Hove \cite{Po:74} postulated that only
gluons are stopped in high energy hadron--hadron interactions and
their energy is used for particle production.
The valence quarks are assumed to fly through.
This picture was converted later into a 3--fluid hydrodynamical model
by Katscher and collaborators \cite{Ka:93}.

This leads us to the generalization of the Landau model \cite{Ga:4}
assuming that only the energy of the produced particles 
(inelastic energy) is
thermalized in the early stage volume.
In the case of A+A collisions a fraction of the inelastic energy
(entropy) can be absorbed by barionic fluids in the late stage
of the expansion. Thus the final state inelastic energy (entropy)
should be corrected for this effect as discussed in Refs. \cite{Ga:4,Ga:5}.

{\bf Entropy Production.}
A crucial assumption which allows us to connect properties of the matter
at the early stage is an assumption that the entropy is produced only
during the very first non--equilibrium stage of the collision,
and it remains constant in the expansion, hadronization and
freeze--out stages.
Isentropic expansion of strongly interacting matter was postulated first
by Landau  \cite{La:53} on the base of qualitative arguments;
at very high energy densities one expects much shorter mean free path
than the size of the system.
The influence of the hadronization on the entropy content depends on 
the nature of the  hadronization process which remains still unclear
\cite{Cs:95, Ra:96}.
Recent studies indicate that entropy seems to be only weakly
affected in the freeze--out stage \cite{So:92,Oc:96}.

{\bf Strangeness Production.}
In the model it is assumed that the strangeness reaches its equilibrium
value at the early stage and remains unchanged during expansion,
hadronization and freeze--out.
Equilibration of strangeness during the high temperature stage of the
expansion in the case of the QGP creation may be expected due to the fact
that the estimated strangeness equilibration time is comparable with the
life time of the QGP \cite{Ra:96}.
Note that in the equilibrium QGP, due to the low mass of the strange quarks,
isentropic expansion implies expansion with approximately constant
strangeness content.
Due to this fact it is not important for the final results at which 
stage of the QGP evolution the equilibration of strangeness takes place.
The validity of the assumption that the strangeness remains constant
in the hadronization stage depends
(as in the case of the entropy) on the nature of this process.
The production of strangeness at the freeze--out stage can be neglected
due to a relatively large mass of strange hadrons and the 
requirement of strangeness
conservation.

\section{Summary and Conclusions}
 
The experimental data on pion and strangeness production indicate:\\
-- saturation of pion and strangeness production with the
number of participant nucleons,\\
-- change in the collision energy dependence taking place between
15 A$\cdot$GeV/c and 160 A$\cdot$GeV/c.\\
Within a  statistical approach
the observed behaviour can be qualitatively understood as due to:\\
-- equilibration of entropy  and strangeness  in
collisions of heavy nuclei,\\
-- transition to a QGP occuring between
AGS and  SPS energies 
associated with the increase of the effective number of degrees of freedom.\\
\noindent
These observations  already hold for central S+S collisions,
they are not unique to central Pb+Pb collisions.

A non--monotonic collision energy dependence of the {\it strangeness/pion}
ratio is expected  in the transition energy region.

The results at SPS energies are in surprisingly good agreement 
with the calculations done within the  the
generalized Landau model.
The analysis suggests that the plasma created at the 
SPS may have a temperature of
about 280 MeV (energy density of about 10 GeV/fm$^3$).

Experimental studies of central Pb+Pb collisions in the energy range
20--160 A$\cdot$GeV are {\bf urgently} needed in order to localize 
the threshold energy more precisely and to study the properties
of the QCD phase transition.

{\it Acknowledgements.}  
I would like to thank Christian Bormann for the comments to the
manuscript.

\smallskip

%
 
\newpage

\begin{figure}[t]
\epsfig{file=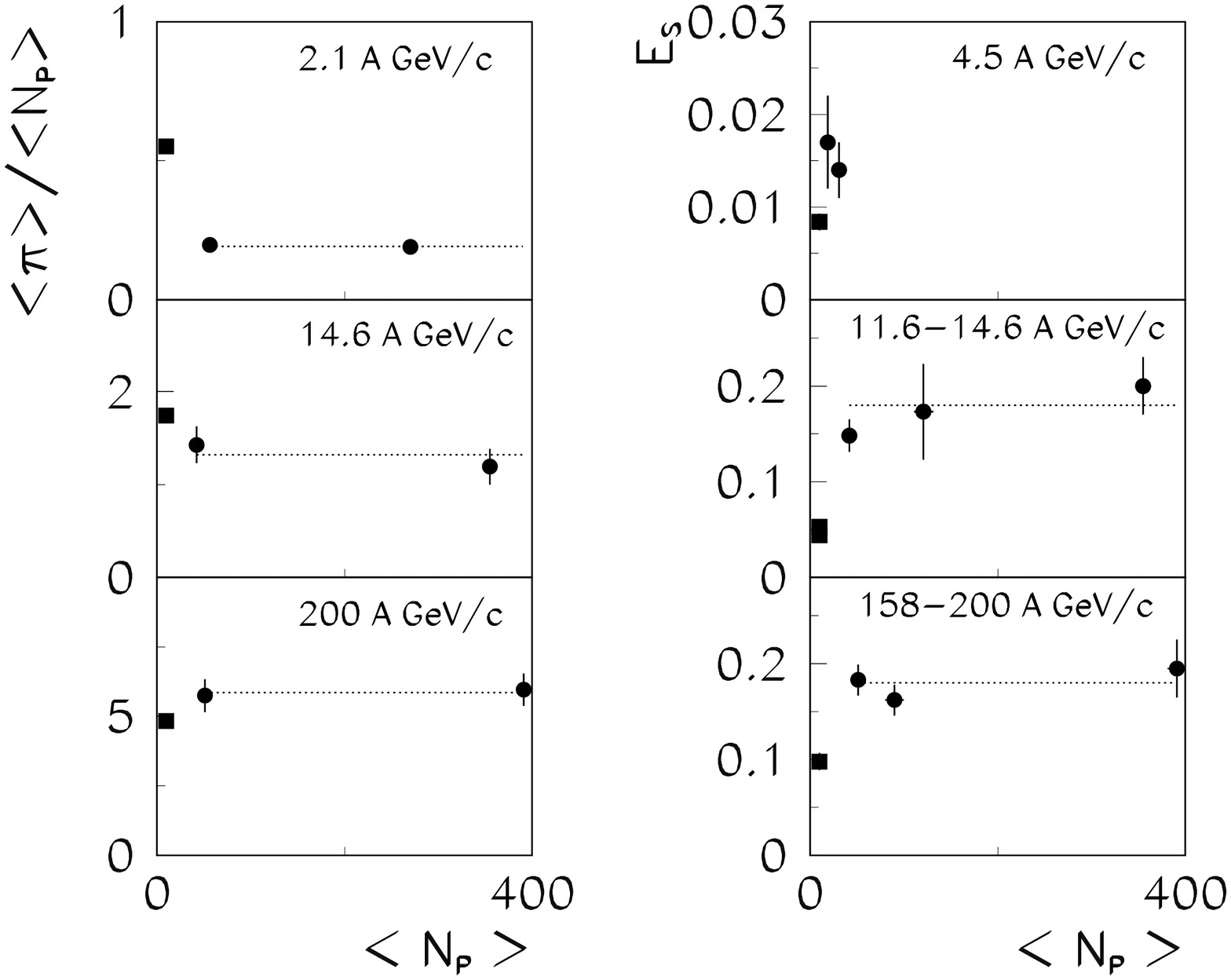,height=12cm}
\caption{ \protect\begin{small}
(left)
The dependence of the ratio $\langle \pi \rangle/\langle N_P \rangle$
on $\langle N_P \rangle$ at three different collision energies.
(right)
The dependence of the $E_S$ ratio
on $\langle N_P \rangle$ at three different collision energies.
The data for central A+A collisions are indicated by circles and
the data for N+N interactions by squares.
\protect\end{small} }
\label{fig1}
\end{figure}

\newpage

\begin{figure}[t]
\epsfig{file=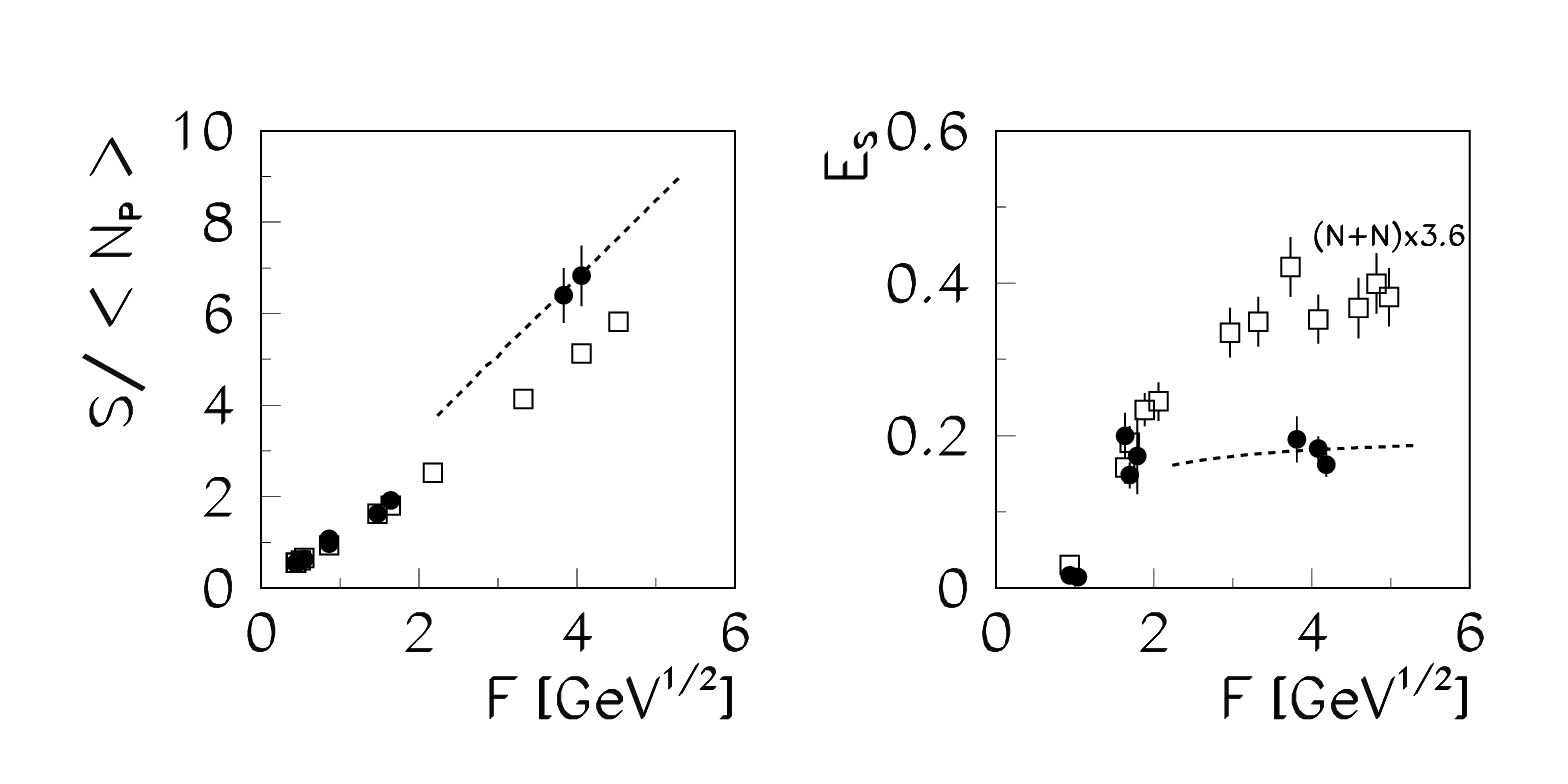,height=7cm}
\caption{\protect\begin{small}
The dependence of  $S/\langle N_P \rangle$ (left) and $E_S$ (right)
on the collision energy measured by the Fermi variable $F$.
The data for central A+A collisions are indicated by circles and the
data for N+N interactions by squares (the $E_S$ values for N+N interactions are
scaled by a factor of 3.6).
The dashed lines show results obtained within the generalized Landau model
assuming QGP creation.
\protect\end{small}}
\label{fig2}
\end{figure}


\begin{thebibliography}{99}
\bibitem{St:96} H. Str\"obele, in Proceedings of Quark Matter 96,
note that the values of inelasticity for S+S and Pb+Pb collisions 
at SPS are plotted
incorrectly~\cite{Ba:94}.
\end{thebibliography}
\end{document}